\documentstyle[aps,12pt]{revtex}
\begin{document}
\draft
\author{O. B. Zaslavskii}
\address{Department of Physics, Kharkov Karazin's National University, Svoboda Sq.4,\\
Kharkov\\
61077, Ukraine\\
E-mail: aptm@kharkov.ua}
\title{Entropy of massive fields near a black hole and vacuum polarization:
thermodynamics without statistical mechanics}
\maketitle

\begin{abstract}
Starting from the Frolov-Zel'nikov stress-energy tensor of quantum massive
fields in the Schwarzschild background, we recover the contribution $S_{q}$
of these field into the entropy of a black hole. For fermions with the spin $%
s=1/2$ $S_{q}>0$, for scalar fields $S_{q}>0$ provided the coupling
parameter is restricted to some interval, and $S_{q}<0$ for vector fields.
The appearance of negative values of $S_{q}$ is attributed to the fact that
in the situation under discussion there are no real quanta to contribute to
the entropy, so $S_{q}$ is due to vacuum polarization entirely and has
nothing to do with the statistical-mechanical entropy. We also consider the
spacetime with an acceleration horizon - the Bertotti-Robison spacetime -
and show that $S_{q}=0$ for massive fields similarly to what was proved
earlier for massless fields.
\end{abstract}


%\date{\today}

\section{introduction}

One of the most striking features in black hole physics consists in that a
black hole behaves like a thermodynamic system and possesses temperature and
entropy. In the state of thermal equilibrium the total entropy $%
S_{tot}=S_{bh}+S_{q}$. Here S$_{bh}=A/4$ ($%TCIMACRO{\UNICODE[m]{0x127}}
%BeginExpansion
\rlap{\protect\rule[1.1ex]{.325em}{.1ex}}h%
%EndExpansion
=c=1$) is the Bekenstein-Hawking entropy ($A$ is the surface area of the
event horizon), $S_{q}$ is the contribution of radiation and matter fields.
The presence of quantum fields modifies the entropy and gives rise to a
number of nontrivial issues. The most popular one is the problem of quantum
renormalization of the entropy and removing thermal divergencies due to
fluctuations of fields propagating in the vicinity of the horizon (see, e.g.
the book \cite{fn}). Less attention is paid to the final result of the
regularization and its properties, i.e. $S_{q}$ which enters all
thermodynamics relations and describes concrete properties of a system.
Thermodynamics of a black hole dressed by Hawking radiation was considered
in a series of papers \cite{mat} but, to the best of my knowledge, for
massless fields only.

Some general remarks on possible nontrivial consequences for thermodynamics
of a black hole with a strong backreaction of massive fields were suggested
in \cite{notes} but without discussing the issue of the entropy. It would
seem that thermodynamics of massive fields is less interesting than that of
massless ones since if a mass of fields is large enough, $m\gg T$, where $T$
is the Hawking temperature, the contribution of thermal excitations into all
thermodynamic quantities including entropy, is negligible. For an
''ordinary'' (without a black hole, say, relativistic star) system this
means that all thermal contributions to relevant thermodynamic quantities
(energy, free energy, entropy, etc.) are exponentially small. There is,
however, a big difference between the behavior of the energy and entropy of
such a system. The energy in the limit under consideration tends to nonzero
quantity that arises due to effects of vacuum polarization and depends on
details of geometry. In contrast to it, the entropy in the limit under
consideration remains exponentially small, i.e. negligible.

The situation changes drastically for the case of a black hole. It is
essential that the contribution of vacuum polarization and true particles
into the stress-energy tensor can be separated for massive fields, the first
term being dominant \cite{fn}, \cite{fz}. We exploit this observation for
thermodynamics to argue that even in the limit $m\gg T$ we are left with
some nonzero contribution into the thermodynamic entropy depending on the
geometry of the background and induced by vacuum polarization (zero
oscillations) of quantum fields. In fact, in the main approximation the
thermodynamic entropy of quantum fields is due to vacuum polarization
entirely. Thus, it is the condition $m\gg T$ which reveals the qualitative
difference in the thermodynamics of systems with and without black holes.
Moreover, it turns out that the contribution under discussion may have any
sign (thus, $S_{q}<0$ for the spin $s=1)$.

All what was said above can serve as a motivation to the calculation of $%
S_{q}$ for massive fields. To solve this problem, we apply, {\it mutatis
mutandis}, the approach elaborated in previous papers \cite{rad1}, \cite
{rad2} where it was used for the massless case.

\section{Entropy in the Schwarzschild background}

Consider a quantum field $\phi $ at finite temperature equal to its Hawking
value $T=\beta ^{-1}$. Its Euclidean action takes the standard thermodynamic
form 
\begin{equation}
I_{q}=-\beta \int d^{3}x\sqrt{g}T_{0}^{0}-S_{q}\text{.}  \label{action}
\end{equation}
Here $T_{\mu }^{\nu }$ is the stress-energy tensor of quantum fields
calculated with respect to the background with the metric $g_{\mu \nu }$, $g$
is determinant of $g_{\mu \nu }$. It is assumed from the very beginning that
the renormalization is already performed in $T_{\mu }^{\nu }$, so we may
manipulate with $T_{\mu }^{\nu }$ safely. We consider the Schwarzschild
background: 
\begin{equation}
ds^{2}=-dt^{2}(1-r_{+}/r)+(1-r_{+}/r)^{-1}dr^{2}+r^{2}d\omega ^{2}
\label{metric}
\end{equation}

The system is supposed to be confined in the spherical container of the
radius $R$. Consider now the response of the action to the particular class
of metric variations caused by the change of the value of $r_{+}$ which
preserve the general form (\ref{metric}). Then, as was shown in (\cite{rad2}%
), 
\begin{equation}
(4\pi )^{-2}\frac{\partial I_{q}}{\partial r_{+}}=R^{3}T_{r}^{r}(R)-%
\int_{r_{+}}^{R}drr^{2}T_{\mu }^{\mu }  \label{der}
\end{equation}

The stress-energy of massive fields for $m\gg T$ has the general structure 
\cite{fz}, \cite{fn} 
\begin{eqnarray}
T_{\mu }^{\nu } &=&\frac{r_{+^{2}}}{r^{8}}f_{\nu }^{\mu }(\frac{r_{+}}{r})%
\text{, }  \label{tensor} \\
f_{\nu }^{\mu }(\frac{r_{+}}{r}) &=&\frac{A}{m^{2}}q_{\mu }^{\nu }\text{, }%
q_{\mu }^{\nu }=a_{\mu }^{\nu }+b_{\mu }^{\nu }\frac{r_{+}}{r}\text{, } 
\nonumber
\end{eqnarray}
where $A$, $a_{\mu }^{\nu }$ and $b_{\mu }^{\nu }$ are pure numbers.

It is convenient to introduce a variable $u=r_{+}/r$, $w\leq u\leq 1$, $%
w=r_{+}/R$. Then we have the equation 
\begin{equation}
\frac{\partial J}{\partial w}=w^{2}f_{1}^{1}(w)+w^{-3}\int_{1}^{w}duu^{4}f_{%
\mu }^{\mu }(u)\text{,}  \label{w}
\end{equation}
where $J\equiv (4\pi )^{2}R^{2}I_{q}$. Taking into account the identity 
\begin{eqnarray}
\int_{1}^{w}du^{\prime }f(u^{\prime })\int_{1}^{u^{\prime }}du^{\prime
\prime }g(u^{\prime \prime }) &=&\int_{1}^{w}dug(u)[\psi (w)-\psi (u)]\text{,%
}  \label{iden} \\
\frac{d\psi (u)}{du} &\equiv &f(u)\text{,}  \nonumber
\end{eqnarray}
and the relationship between the action and entropy (\ref{action}), we get
after simple manipulations: 
\begin{eqnarray}
S_{q} &=&\frac{16\pi ^{2}}{m^{2}r_{+}^{2}}A\Phi (w)\text{,}  \label{ent} \\
\Phi (w) &=&w^{2}\int_{w}^{1}duu^{2}\chi _{1}(u)+\int_{w}^{1}du\chi _{2}(u)%
\text{,}  \nonumber \\
\chi _{1} &=&q_{1}^{1}+\frac{q_{\mu }^{\mu }}{2}\text{, }\chi
_{2}=-u^{2}(q_{0}^{0}+\frac{q_{\mu }^{\mu }}{_{2}})\text{.}  \nonumber
\end{eqnarray}

The constant of integration is chosen in such a way that $S_{q}=0$ at $w=1$, 
$r_{+}=R$ (no room for radiation).

It follows from (\ref{ent}) and from the regularity condition on the horizon 
$q_{0}^{0}(1)=q_{1}^{1}(1)$, equivalent to $%
T_{0}^{0}(r_{+})=T_{1}^{1}(r_{+}) $, that $\frac{d\Phi }{dw}_{\mid w=1}=0$,
so $\Phi (w)=(1-w)^{2}\rho $, where $\rho $ is some function of $w$, finite
at $w=1$. Its explicit form will be given below for different types of
fields.

Restoring the usual variables, we can rewrite the contribution of quantum
fields in the form 
\begin{equation}
S_{q}=\beta _{H}\int d^{3}x\sqrt{g}[\frac{r^{2}}{R^{2}}(T_{1}^{1}+\frac{1}{2}%
T_{\mu }^{\mu })-(T_{0}^{0}+\frac{1}{2}T_{\mu }^{\mu })]  \label{fin}
\end{equation}
or 
\begin{eqnarray}
S_{q} &=&\int d^{3}x\sqrt{g_{3}}s_{ef}\text{,}  \label{euler} \\
T_{loc}s_{ef} &=&p+\rho -(1-\frac{r^{2}}{R^{2}})(\rho +\frac{1}{2}T_{\mu
}^{\mu })\text{,}  \nonumber
\end{eqnarray}
where $p\equiv T_{1}^{1}$ is the radial pressure, $\rho =-T_{0}^{0}$ is the
energy density, $T_{loc}=T/\sqrt{-g_{oo}}$ is a local Tolman temperature, $%
g_{3}$ is the determinant of the spatial metric. Eq. (\ref{euler}) replaces
the usual thermodynamic Euler relation $T_{loc}s=P+\rho $.

The method of the derivation of the expression for the entropy may be
applied not only to the quantity $S_{q}$ as a whole but also to the
distribution of the entropy in space. It is obvious in our approach that for
any layer $r_{+}\leq r\leq r_{0}\leq R$ one obtains the same formulas (\ref
{ent})-(\ref{euler}) with the replacement $R\rightarrow r_{0}$.

Performing integration in (\ref{ent}) and using the explicit values of the
coefficients $a_{\mu }^{\nu }$ and $b_{\mu }^{\nu }$ according to \cite{fn}
and \cite{fz}, we obtain after simple calculations (the index indicates the
value of the spin of a field): 
\begin{eqnarray}
\Phi _{s} &=&(1-w)^{2}\rho _{s}\text{,}  \label{f} \\
\rho _{0} &=&(122-546\xi )w^{4}+(94-420\xi )w^{3}+(66-294\xi
)w^{2}+(38-168\xi )w+19-84\xi \text{,}  \label{0} \\
\rho _{1/2} &=&8+16w+30w^{2}+44w^{3}+58w^{4}\text{,}  \label{ferm} \\
\rho _{1} &=&-3(9+18w+32w^{2}+46w^{3}+60w^{4})\text{.}  \label{1}
\end{eqnarray}

\section{discussion}

Let us now discuss the obtained results. In the spirit of (\cite{mat}) for
the scalar field we could derive from the conditions of the absence of
layers with the negative entropy $\Phi \geq 0$ and $\frac{d\Phi }{dw}<0$
(equivalent to ($\frac{\partial S_{q}}{\partial R}$)$_{r_{+}}>0$), the
restrictions on the coupling parameter $\xi _{1}\leq \xi \leq \xi _{2}$,
where in our case $\xi _{1}=$ $0.2142858$ and $\xi _{2}=0.2242068$. Thus,
both the conformal ($\xi =1/6$) and minimal ($\xi =0$) couplings are outside
this range. However, the example with the vector field especially clearly
shows that the values $S_{q}<0$ are admissible, so restriction on the
coupling parameter in the scalar case can be relaxed at all.

As the result $S_{q}<0$ looks rather surprising, we would like to stress the
following. First, the expression for the tensor (\ref{tensor}) was derived
under the assumption that $m\gg T$. It means that the contribution of
thermal excitations to the entropy and other thermodynamic quantities are
strongly damped with the exponential factor $\exp (-m/T)$ and one deals with
the vacuum polarization effects entirely. In the limit $R\rightarrow \infty $
$T_{\mu }^{\nu }\rightarrow 0$ in contrast to the massless case when the
tensor $T_{\mu }^{\nu }$ approaches that of thermal radiation in the flat
space. The contribution from the statistical-mechanical part $S^{SM}$ (that
is the sum over modes with the Bose distribution function) is absent in the
main approximation - correspondingly, any reasonable statistical
interpretation of $S_{q}$ cannot be given and this abolishes the prohibiting
of S$_{q}<0$.

Second, the quantity $S_{q}$ enters the expression for the total entropy of
a black hole with its counterpart $S_{bh\text{ }}$only: $%
S_{tot}=S_{bh}+S_{q} $. As $S_{q}$, by its very meaning, represents small
corrections to $S_{bh}$, it is obvious that $S_{tot}>0$. The term $S_{q}$
arises due to the presence of a black hole and cannot be ''cut'' from it and
considered on its own. For comparison, in the vicinity of a relativistic
star there is no $S_{bh}$, but $S_{q}$ is due to thermal excitations only
and is positive (exponentially small, if $m\gg T$).

Usually, the contribution to the total thermodynamic entropy of a black hole
which comes from quantum fields can be represented in the form 
\begin{equation}
S_{q}\equiv S^{TM}=S^{SM}+S_{0}\text{,}  \label{mech}
\end{equation}
where $S^{SM}$ is the statistical-mechanical part of the entropy and $S_{0}$
is some qiantity depending on parameters of the geometry (say, the horizon
radius in the Schwarzschild case). Both $S^{SM}$ and $S_{0}$ diverge due to
the contribution of the near horizon region but their sum is finite. This
scheme explains the general mechanism of the black hole entropy
renormalization \cite{fn}, \cite{frolov}, \cite{zurthorn}. The quantity $%
S^{TM}$ represents the thermodynamic entropy, i.e. the entity that enters
the general thermodynamic relations - in particular, it describes the
response of the free energy to the change of temperature \cite{xren}.

The representation (\ref{mech}) was directly justified in the terms of the
stress-energy tensor for massless fields \cite{rad2}. In that case 
\begin{equation}
S_{q}=16\pi ^{2}r_{+}\int_{r_{+}}^{R}drr^{2}(T_{1}^{1}-T_{0}^{0}-T_{\mu
}^{\mu }\ln \frac{R}{r})\text{.}  \label{sch}
\end{equation}
Using the splitting for the stress-energy tensor of quantum fields in the
Hartle-Hawking state \cite{split} 
\begin{equation}
T_{\mu }^{\nu }=(T_{\mu }^{\nu })_{B}+(T_{\mu }^{\nu })_{th}  \label{split}
\end{equation}
(index ''B'' denote the Boulware state, ''th'' stands for thermal radiation)
we obtain just (\ref{mech}) with $S^{SM}$ identified with the entropy $%
S^{th} $ of thermal radiation and $S_{0}$ obtained by the replacement of $%
T_{\mu }^{\nu }$ in (\ref{sch}) by $(T_{\mu }^{\nu })_{B}$.

For the case of massive fields the situation is quite different. Now $%
S^{SM}=0$ in the main approximation with respect to the parameter $T/m$,
while $S_{0}$ is finite and should be identified (up to small corrections)
with $S_{q}$.

The fact that $S^{SM}$ is discarded has important consequences for the
scheme of a black hole renormalization in the case of massive fields. The
detailed discussion of this issues is beyond the scope of the present paper,
so we will restrict ourselves by several remarks. In the paper \cite{brick}
it was observed that using Pauli-Villars regularization enables one to
obtain the finite value of the entropy without introducing a cutoff
parameter typical for the brick wall model. The approach of \cite{brick}
well explains the mechanism of cancelling divergencies but, in our view,
cannot give the correct value for the finite part of $S_{q}$ describing the
contribution of matter fields that depends on the position of the shell
surrounding a black hole. (And, in fact, the authors of \cite{brick} did not
pose the purpose of full calculating such a quantity and themselves
emphasize that they neglect contributions to the entropy that does not
diverge as the mass of the regulator fields tends to infinity. Meanwhile,
this contribution is essential for calculating the entropy contained between
the horizon and the shell, i.e. the quantity we are seeking for.) This is
due to the fact that $S_{q}$ is represented in \cite{brick} as the linear
combination of quantities $S^{SM}$ for different kinds of fields, while the
contribution of corresponding terms $S_{0}$ was not taken into account.
Meanwhile, as we emphasize, $S^{SM}=0$ in the main approximation at hand,
the main contribution to $S_{q}$ comes from the vacuum polarization, so
account for $S_{0}$ (not made in \cite{brick}) is crucial for our task.

It is worth stressing that, although $S_{q}$ cannot be given statistical
interpretation, it does contribute to thermodynamic relations (for instance,
into the general first law), so we have ''thermodynamics without statistical
mechanics''. As it is statistical interpretation that demands $S_{q}>0$, the
lack of such an interpretation opens the possibility for $S_{q}<0$.

\section{Bertotti-Robinson spacetime}

Discussion above refers to the black hole background. Now we will consider
another type of spacetime with a horizon - Bertotti-Robinson spacetime (BR)
that, as is known, possesses the acceleration horizon \cite{br}, \cite{lap}.
Recently it was shown that in such a spacetime $S_{q}=0$ for massless fields 
\cite{ent} (generalization to other spacetimes with an acceleration horizon
was suggested in \cite{00}). This fact was attributed to the pure kinematic
nature of an acceleration horizon which does not produce ''true'' quanta.
This suggests that the result $S_{q}=0$ must hold true for massive fields on
the same footing as in the massless case. Below we confirm it by direct
calculation.

The variation of the action with respect to the metric in the BR has one
subtlety: as the radius $r_{+}$ enters the expression for the angular part
of the metric on a boundary, the components $g_{\theta \theta }$ and $%
g_{\phi \phi }$ vary along with components $g_{00}$ and $g_{rr}$. In
particular, in the formula for the first general law this gives rise to
additional terms responsible for the gravitational contribution of the
pressure. These points are discussed in details in \cite{00}, so below we
only list the corresponding results without derivation.

The metric under discussion has the form 
\begin{eqnarray}
ds^{2} &=&-dt^{2}b^{2}+dl^{2}+r_{+}^{2}d\omega ^{2}\text{,}  \label{br} \\
b &=&r_{+}d(z)\text{, }d=shz\text{, }z=l/r_{+}\text{.}  \nonumber
\end{eqnarray}

In this background the stress-energy of massive fields reads \cite{sah}, 
\cite{jur} 
\begin{equation}
T_{\mu }^{\nu }=\frac{C}{r_{+}^{6}}(1,1,-1,-1)\text{.}  \label{strbr}
\end{equation}
Here $C=c/m^{2}$, $c$ is a numerical constant whose exact value is
irrelevant for our purposes. From dimensional arguments and the structure of 
$T_{\mu }^{\nu }$ it follows that the semiclassical action of a massive
quantum field confined in the region $0\leq z\leq z_{B}$%
\begin{equation}
I_{q}=\frac{J(z_{B})}{r_{+}^{2}}\text{.}  \label{brac}
\end{equation}
Now let us take advantage of the relation (see eq. (10) of \cite{00}): 
\begin{equation}
(\frac{\partial I}{\partial z})_{r_{B}}=-8\pi ^{2}r_{+}^{2}d(z)T_{1}^{1}%
\text{.}  \label{10}
\end{equation}
Then we obtain that 
\begin{equation}
J=-8\pi C\int_{0}^{z_{B}}dz^{\prime }d(z^{\prime })\text{,}  \label{j}
\end{equation}
where we took into account that $J=0$ if $z_{B}=0$ (no room for quantum
fields). Comparing (\ref{brac}), (\ref{j}) with (\ref{action}), we get
immediately $S_{q}=0$.

This result coincides with that in \cite{ent}, \cite{00} and can be
explained by the same reason. It consists in that the horizon of the BR is
not a true black hole horizon but represents a pure kinematic effect, too
weak to gain nonzero entropy. However, there is also a difference between
interpretation of the property $S_{q}=0$ for massless and massive fields. In
the first case this property can be thought of as the result of the mutual
compensation between two divergent contributions ($S^{SM}$ and $S_{0}$) from
thermal excitations and vacuum polarization. For the massive field case the
first contribution is absent, so we have $S_{0}=0$.

\section{conclusion}

To the best of my knowledge, the situation we discussed is the first example
when the thermodynamic entropy arises due to vacuum polarization effects, so
the entropy reveals itself even in spite of the absence of thermal
excitations. This phenomenon is due to the presence of an event horizon and
is absent, for instance, in the background of a relativistic star. In the
latter case quantum-gravitational polarization effects cannot themselves
produce entropy that may arise due to thermal excitations of quantum fields
only.

The unusual situation, called above ''thermodynamics without statistical
mechanics'', reveals itself, in particular, in that values of corrections $%
S_{q}<0$ are allowed. In spite of the difference between two examples
considered in the present paper they share the common feature - the absence
of ''true'' quanta that resulted in the unusual properties $S_{q}<0$
(Schwarzschild background) and $S_{q}=0$ (BR). It would be of interest to
verify the properties outlined above in other black hole spacetimes.

In conclusion, let me point out the important issues that remained beyond
the scope of the present paper and need separate treatment. It would be
interesting to present microscopic description of results obtained and
understand, how they can be interpreted in terms of counting quantum states.
This problem becomes especially interesting in the context of recent
developments in the framework of the D-brane approach.

\section{acknowledgment}

I\ thank H. Nicolai and Albert Einstein Institute, where this work was
performed, for hospitality. I\ am grateful to S. N. Solodukhin for reading
the manuscript and discussion.


% insert suggested PACS	numbers	in braces on next line

% body of paper	here

% now the references. delete or	change fake bibitem. delete next three
%   lines and directly read in your .bbl file if you use bibtex.

% Here is an example of	the general form of a figure:
% that you will	use with \ref{}	command	in the braces of the \label{} command.
%
% \begin{figure}
% \caption{}
% \label{}
% \end{figure}

% tables follow	here
%
% Here is an example of	the general form of a table:
% that you will	use with \ref{}	command	in the braces of the \label{} command.
% \begin{table}
% \caption{}
% \label{}
% \end{tabular}
% \end{table}


\begin{references}
\bibitem{fn}  V.P. Frolov and I.D. Novikov, Black Hole Physics (Kluwer
Academic Publishers, vol. 1996, Chapter 11.1.3.7).

\bibitem{mat}  D. Hochberg, T.W. Kephart and J.W. York, Jr, Phys. Rev. D 48
(1993) 479.

P.R. Anderson, W.A. Hiscock, J. Whitesell and J.W. York, Jr, Phys. Rev{\it . 
}D 50 (1994) 6427.

D. Hochberg and S.V. Sushkov, Phys. Rev{\it . }D 53 (1996) 7094.

\bibitem{notes}  O.B. Zaslavskii, Class. Quant. Grav. 8 (1991) L141-L145.

\bibitem{fz}  V.P. Frolov and A.I. Zel'nikov, Phys. Lett. B 115 (1982) 372.

\bibitem{rad1}  O.B. Zaslavskii, Phys. Lett{\it . }A 181 (1993 )105.

\bibitem{rad2}  O.B. Zaslavskii, Class. Quant. Grav. 13 (1996 ) L23.

\bibitem{frolov}  V.P. Frolov Phys. Rev. Lett. 74 (1995 {\bf ) }3319.

\bibitem{zurthorn}  W.H. Zurek and K.S. Thorne, Phys. Rev. Lett. 54 (1985)
2171.

\bibitem{xren}  For non-minimaly coupled matter the mechanism of
renormalization of the black hole entropy has some additional subtleties
(see S. N. Solodukhin, Phys. Rev. D 56 (1997) 4968) but we do not discuss
this issue here.

\bibitem{split}  S. M. Cristensen and S.A. Fulling, Phys. Rev. D 15 (1987)
2089; P.Candelas, Phys. Rev. D 21 (1980) 2185; V.P. Frolov and K.S. Thorne,
Phys. Rev. D 39 (1989) 2125.

\bibitem{brick}  J-G. Demers, R. Lafrance and R. Myers, Phys. Rev. D 52
(1995) 2245.

\bibitem{br}  Robinson I Bull.Acad.Pol.Sci{\it .} 7 (1959) 351.

Bertotti B Phys. Rev. 116 (1959 ) 1331.

\bibitem{lap}  B. Carter General theory of stationary black hole states. In: 
{\it Black Holes}, edited by De Witt C and De Witt B S (New York: Gordon and
Breach, 1973).

\bibitem{ent}  O.B. Zaslavskii, Phys. Rev.\ D 57 (1998 ) 6265.

\bibitem{00}  O.B. Zaslavskii, Class. Quant. Grav. 17 (2000) 497-512.

\bibitem{sah}  L.A. Kofman and V. Sahni, Phys. Lett. B 127 (1983) 197.

\bibitem{jur}  J. Matyjasek, Phys. Rev. D 61, 124019 (2000).
\end{references}
\end{document}